\documentclass[prc,preprint,showpacs,showkeys,nofootinbib]{revtex4}%
\usepackage{amssymb}
\usepackage{graphicx}
\usepackage{bm}
\usepackage{amsmath}
\usepackage{amsfonts}%
\begin{document}
\title{Coulomb excitation of unstable nuclei at intermediate energies}
\author{C.A.~Bertulani$^{1}$\footnote{bertulanica@ornl.gov.},
G. ~Cardella$^{2}$, M. De Napoli$^{2,3}$, G. Raciti$^{2,3}$, and
E. Rapisarda$^{2,3}$}
\affiliation{ $^{1}$ Department of Physics and Astronomy,
University of Tennessee, Knoxville, Tennessee 37996, USA \\
$^{2}$
Istituto Nazionale di Fisica Nucleare, Sezione di Catania, via Santa
Sofia 64, I-95123, Catania,
Italy\\
$^{3}$ Dipartimento di Fisica e Astronomia, Universit\'a Catania,
via Santa Sofia 64, I-95123, Catania, Italy}

\begin{abstract}
We investigate the Coulomb excitation of low-lying states of
unstable nuclei in intermediate energy collisions
($E_{lab}\sim10-500$ MeV/nucleon). It is shown that the cross
sections for the $E1$ and $E2$ transitions are larger at lower
energies, much less than 10 MeV/nucleon. Retardation effects and
Coulomb distortion are found to be both relevant for energies as low
as 10 MeV/nucleon and as high as 500 MeV/nucleon. Implications for
studies at radioactive beam facilities are discussed.

\end{abstract}
\pacs{25.60.-t, 25.70.-z, 25.70.De} \keywords{Coulomb excitation,
cross sections, unstable nuclei.} \maketitle
\date{\today}

Unstable nuclei are often studied with reactions induced by
secondary radioactive beams. Examples of these reactions are elastic
scattering, fragmentation and Coulomb excitation by heavy targets.
Coulomb excitation is specially useful since the interaction
mechanism is very well known \cite{AW75}. It is the result of
electromagnetic interactions of a projectile ($Z_{P}$,$A_{P}$) with
a target ($Z_{T}$,$A_{T}$). One of the participating nuclei is
excited as it passes through the electromagnetic field of the other.
Here we will only consider the excitation of the projectile as is of
interest in studies carried out in heavy ion facilities around the
world, e.g. LNS/Catania, NSCL/MSU, GSI, GANIL, RIKEN, etc. In
Coulomb excitation a virtual photon with energy $E$ is absorbed by
the projectile. Because in pure Coulomb excitation the participating
nuclei stay outside the range of the nuclear strong force, the
excitation cross section can be expressed in terms of the same
multipole matrix elements that characterize excited-state gamma-ray
decay, or the reduced transition probabilities,
$B(\pi\lambda;J_{i}\rightarrow J_{f})$. Hence, Coulomb excitation
amplitudes are strongly coupled with valuable nuclear structure
information. Therefore, this mechanism has been used for many years
to study the electromagnetic properties of low-lying nuclear
states~\cite{AW75}.

Coulomb excitation cross sections are large if the adiabacity parameter
satisfies the condition
\begin{equation}
\xi=\omega_{fi}\dfrac{a_{0}}{v}<1\ , \label{1.1}%
\end{equation}
where $a_{0}$ is half \ the distance of closest approach in a
head-on collision for a projectile velocity $v$, and
$E_{x}=\hbar\omega_{fi}$ is the excitation energy. This adiabatic
cut-off limits the possible excitation energies below 1-2 MeV in
sub-barrier collisions. A possible way to overcome this limitation,
and to excite high-lying states, is to use higher projectile
energies. In this case, the closest approach distance, at which the
nuclei still interact only electromagnetically, is of order of the
sum of the nuclear radii, $R=R_{P}+R_{T}$, where $P$ refers to the
projectile and $T$ to the target. For very high energies one has
also to take into account the Lorentz contraction
of the interaction time by means of the Lorentz factor $\gamma=(1-v^{2}%
/c^{2})^{-1/2}$, with $c$ being the speed of light. For such collisions the
adiabacity condition, Eq.~(\ref{1.1}), becomes
\begin{equation}
\xi(R)=\frac{\omega_{fi}R}{\gamma v}<1\ . \label{1.3}%
\end{equation}
From this relation one obtains that for bombarding energies around
and above 100 MeV/nucleon, states with energy up to 10-20 MeV can be
readily excited \cite{WA79}.

An appropriate description of Coulomb excitation at intermediate
energies ($E_{lab}=10-500$ MeV/nucleon) has been described in ref.
\cite{Ber03}. In this energy region neither the non-relativistic
Coulomb excitation formalism described in ref. \cite{AW75}, nor the
relativistic one formulated in refs. \cite{WA79,BB85} are
appropriate. This is discussed in details in ref. \cite{Ber03} where
it is shown that the correct values of the Coulomb excitation cross
sections differ by up to 30-40\% when compared to the
non-relativistic and relativistic treatments used to calculate
experimental observables (cross sections, gamma-ray angular
distributions, etc.).

We follow the formalism of ref. \cite{Ber03} to calculate cross
sections for Coulomb excitation from energies varying from 10 to 500
MeV/nucleon. These are the energies where most radioactive beam
facilities are or will be operating around the world. The calculated
cross sections will be of useful guide for future experiments. We
also compare the accurate calculations with those obtained by using
simple analytical formulas and test the regime of their validity.

The cross sections for the transition $J_{i}\rightarrow J_{f}$ in
the projectile are calculated using the equation  \cite{Ber03}
\begin{equation}
{\frac{d\sigma_{i\rightarrow f}}{d\Omega}}=\frac{4\pi^{2}Z_{T}^{2}e^{2}}%
{\hbar^{2}}\;a^{2}\epsilon^{4}\;\sum_{\pi\lambda\mu}{\dfrac{B(\pi\lambda
,J_{i}\rightarrow J_{f})}{(2\lambda+1)^{3}}}\;\mid S(\pi\lambda,\mu)\mid
^{2}\ , \label{cross_2}%
\end{equation}
where $\pi=E$ or $M$ stands for the electric or magnetic multipolarity, and
\begin{equation}
B(\pi\lambda,J_{i}\longrightarrow J_{f})=\frac{1}{2J_{i}+1}\ \left\vert
\left\langle J_{f}\left\Vert \mathbf{\mathcal{M}}(\pi\lambda)\right\Vert
J_{i}\right\rangle \right\vert ^{2}\ \label{reduced}%
\end{equation}
are the reduced transition probabilities. In these equations, $\epsilon
=1/\sin(\Theta/2)$, with $\Theta$ being the deflection angle, $a_{0}%
={Z_{P}Z_{T}e^{2}/}m_{0}v^{2}$ and $a=a_{0}/\gamma$. The complex
functions $S(\pi\lambda,\mu)$ are integrals along Coulomb
trajectories corrected for retardation. Their calculation and how
they relate to the non-relativistic and relativistic theories are
described in details in ref. \cite{Ber03}. Here we will introduce
another comparison tool for the total cross section, which is
obtained by integration of eq. \ref{cross_2} over scattering angles.
The code COULINT \cite{Ber03} was used to calculate the orbital
integrals $S(\pi\lambda,\mu)$ and the cross sections of eq.
\ref{cross_2} (for more details, see ref. \cite{Ber03}).

Using the theory described in ref. \cite{BB85}, it is easy to show that
approximate values of the cross sections for E1, E2, and M1 transitions can be
obtained by means of the relations%
\begin{align}
\sigma_{E1}^{(app)}  &  =\frac{32\pi^{2}}{9}\frac{Z_{T}^{2}\alpha}{\hbar
c}B\left(  E1\right)  \left(  \frac{c}{v}\right)  ^{2}\left[  \xi K_{0}%
K_{1}-\frac{v^{2}\xi^{2}}{2c^{2}}\left(  K_{1}^{2}-K_{0}^{2}\right)  \right]
\nonumber\\
\sigma_{E2}^{(app)}  &  =\frac{8\pi^{2}}{75}\frac{Z_{T}^{2}\alpha}{\left(
\hbar c\right)  ^{3}}E_{x}^{3}B\left(  E2\right)  \left(  \frac{c}{v}\right)
^{4}\left[  \frac{2}{\gamma^{2}}K_{1}^{2}+\xi\left(  1+\frac{1}{\gamma^{2}%
}\right)  ^{2}K_{0}K_{1}-\frac{v^{4}\xi^{2}}{2c^{4}}\left(  K_{1}^{2}%
-K_{0}^{2}\right)  \right] \nonumber\\
\sigma_{M1}^{(app)}  &  =\frac{32\pi^{2}}{9}\frac{Z_{T}^{2}\alpha}{\hbar
c}B\left(  M1\right)  \left[  \xi K_{0}K_{1}-\frac{\xi^{2}}{2}\left(
K_{1}^{2}-K_{0}^{2}\right)  \right]  , \label{approx}%
\end{align}
where $K_{n}$ are the modified Bessel functions of the second order, as a
function of $\xi$ given by eq. \ref{1.3}, with $R$ corrected for recoil by the
modification $R\rightarrow R+\pi a/2$ \cite{WA79}.

Here we will only consider the excitation of the lowest lying states
in light and medium heavy nuclei. For nuclear masses $A<20$, the
TUNL nuclear data evaluation web site was of great help \cite{tunl}.
The electromagnetic transition rates at the TUNL database are given
in Weisskopf units and are transformed to the appropriate
$B(\pi\lambda,I_{i}\rightarrow I_{f})$-values by means of the
standard Weisskopf relations $B_W(E1;J_{i}\rightarrow
J_{gs})=0.06446A^{2/3}$
e$^{2}$fm$^{2}$, $B_W(E2;J_{i}\rightarrow J_{gs})=0.05940A^{4/3}$ e$^{2}$%
fm$^{4}$, and $B_W(M1;J_{i}\rightarrow J_{gs})=1.79\left(  e\hbar/2m_{n}%
c\right)  ^{2}$. For comparison, a few medium mass nuclei, as well
as a few stable nuclei, were included in the calculation. Other data
were taken from refs. \cite{Mot95,Sch96,Gad03,Im04}.

Some cases of nuclei far from the stability line are very
interesting and deserve further study, possibly using the method of
Coulomb excitation. For example, it is well known that nuclei with
open shells tend to have $B(E2)$ values greater than 10 W.u.,
whereas nuclei with shell closure of neutrons or protons tend to
have distinctly smaller $B(E2)$ values. Typical examples of the
latter category are the doubly magic nuclei, $^{16}$O and $^{48}$Ca,
which $B(E2)$ values are 3.17 and 1.58 W.u., respectively. According
to an empirical formula adjusted to a global fit of the known
transition rates, the
values of first excited $2^{+}$ level, $E_{2^{+}}$, and $B(E2;0^{+}%
\rightarrow2^{+})$ are related by \cite{Ra88} ($E_{2^+}$ in keV)
\begin{equation}
B(E2;0^{+}\rightarrow2^{+})=26\frac{Z^{2}}{A^{2/3}E_{2^{+}}}\text{ e}%
^{2}\text{fm}^{4}.
\end{equation}

The value of $B(E2)$ for $^{16}$C based on this formula is at least
one order of magnitude larger than what is observed experimentally
in a Coulomb dissociation experiment \cite{Im04}. The anomalously
strong hindrance of the $^{16}$C transition is not well explained
theoretically. This is just an example of the power of Coulomb
excitation as a tool to access the new physics inherent of poorly
known rare nuclear species.

Another example is the strong $E1$ transition in $^{11}$Be.
$^{11}$Be is an archetype of a halo nucleus and exhibits the fastest
known dipole transition between bound states in nuclei. The $B(E1)$
transition strength between the ground and the only bound excited
state (at 0.32 MeV) was determined from lifetime measurements by
Millener et al. to be 0.116 e$^{2}$fm$^{2}$ \cite{Mi83}. However,
Coulomb excitation experiments have obtained a much smaller value of
the $B(E1)$ which is still a matter of investigation
\cite{BCH95,Hus06,Cha07}. It is thus seems clear that predictions based on
traditional nuclear structure and reaction theory often yields
results in disagreement with experimental data. In spite of that,
when proper corrections are accounted for (e.g. channel-coupling,
nuclear excitation, relativistic corrections), Coulomb excitation of
radioactive beams is a powerful complementary tool to investigate
electromagnetic properties of nuclei far from the stability line.

In Table 1 we compare our calculations with several experimentally
obtained cross sections for Coulomb excitation of unstable nuclei.
The units of energy are MeV, the laboratory energy is in
MeV/nucleon, the B-values are in units of e$^{2}$ fm$^{2\lambda}$,
and the cross sections are in millibarns. The last two columns give
the calculated cross sections obtained by using eqs. \ref{cross_2}
and \ref{approx}, respectively. Since the cross sections of  eq.
\ref{approx} are functions of the minimum impact parameter, the
values reported in the Table have been calculated according to the
experimental angular ranges reported in the seventh column. Except
for the $^{11}$Be case, for which the discrepancy between theory and
experiment is known (see discussion above), the calculated cross
sections are close to the experimental values. Nonetheless, the
calculated cross sections tend to be smaller than the experimental
ones for $^{17}$Ne, $^{32}$Mg, $^{38}$S, $^{40}$S, $^{42}$S,
$^{44}$Ar, and $^{46}$Ar projectiles. This is worrisome because the
B($\pi\lambda$) values were extracted from the experimentally
obtained cross sections, using equations similar to eq.
\ref{approx}. These experimental B-values would have to be larger by
$10-30$\% according to our calculations.

It is important to stress the fact that many experimental data on
unstable nuclei collected up to now have been analyzed by means of
theoretical tools (DWBA and coupled-channels codes) which do not
include relativistic {\it dynamics} (the inclusion of relativistic
{\it kinematics} is straightforward). This problem was first
addressed in ref. \cite{Ber05}, where it was shown that the analysis
of experimental data at intermediate energies without a proper
treatment of relativistic dynamics leads to wrong values of
electromagnetic transition probabilities. We should stress that a
full theoretical treatment of relativistic dynamics of strong and
electromagnetic interactions in many-body systems is very difficult
and still does not exist \cite{Ber05}.

\begin{tabular}
[c]{|l|l|l|l|l|l|l|l|l|l|l|}\hline
Data & Projectile & Target & E$_{\text{lab}}$ & $\pi\lambda$ & B($\pi\lambda
$) & $\theta_{\text{range}}$ & E$_{x}$ & $\sigma_{\exp}$ & $\sigma_{\text{th}%
}$ & $\sigma_{\text{app}}$\\\hline\hline 1 \cite{Ann95} & $^{11}$Be
& Pb & 43. & E1 & 0.115 & $<5^{\circ}$ & 0.32
($\frac{1}{2}^{+}\rightarrow\frac{1}{2}^{-}$) & $191\pm26$ & 328. &
323.\\\hline 2 \cite{Ann95} & $^{11}$Be & Pb & 59.4 & E1 & 0.094 &
$<3.8^{\circ}$ & 0.32 ($\frac{1}{2}^{+}\rightarrow\frac{1}{2}^{-}$)
& $304\pm43$ & 213. & 211.\\\hline 3  \cite{Fau97} & $^{11}$Be & Au
& 57.6 & E1 & 0.079 & $<3.8^{\circ}$ & 0.32
($\frac{1}{2}^{+}\rightarrow\frac{1}{2}^{-}$) & $244\pm31$ & 170. &
168.\\\hline 4 \cite{Nak97} & $^{11}$Be & Pb & 64. & E1 & 0.099 &
$<3.8^{\circ}$ & 0.32 ($\frac{1}{2}^{+}\rightarrow\frac{1}{2}^{-}$)
& $302\pm31$ & 217. & 215.\\\hline 5 \cite{Ch97,Ch02} & $^{17}$Ne &
Au & 60. & M1 & 0.163 & $<4.5^{\circ}$ & 1.29
($\frac{1}{2}^{-}\rightarrow\frac{3}{2}^{-}$) & $12\pm4$ & 12.6 &
13.0\\\hline 6 \cite{Mot95} & $^{32}$Mg & Pb & 49.2 & E2 & 454 &
$<4^{\circ}$ & 0.885 ($0^{+}\rightarrow2^{+}$) & $91.7\pm14.4$ &
137. & 128.\\\hline 7 \cite{Ch97} & $^{38}$S & Au & 39.2 & E2 & 235
& $<4.1^{\circ}$ & 1.29 ($0^{+}\rightarrow2^{+}$) & $59\pm7$ & 48. &
45.0\\\hline 8 \cite{Ch97} & $^{40}$S & Au & 39.5 & E2 & 334 &
$<4.1^{\circ}$ & 0.91 ($0^{+}\rightarrow2^{+}$) & $94\pm9$ & 75.5 &
70.4\\\hline 9 \cite{Ch97} & $^{42}$S & Au & 40.6 & E2 & 397 &
$<4.1^{\circ}$ & 0.89 ($0^{+}\rightarrow2^{+}$) & $128\pm19$ & 101.
& 94.3\\\hline 10 \cite{Ch97} & $^{44}$Ar & Au & 33.5 & E2 & 345 &
$<4.1^{\circ}$ & 1.14 ($0^{+}\rightarrow2^{+}$) & $81\pm9$ & 62.3 &
58.3\\\hline 11 \cite{Ch97} & $^{46}$Ar & Au & 35.2 & E2 & 196 &
$<4.1^{\circ}$ & 1.55 ($0^{+}\rightarrow2^{+}$) & $53\pm10$ & 40.9 &
38.2\\\hline 12 \cite{Gad03} & $^{46}$Ar & Au & 76.4 & E2 & 212 &
$<2.9^{\circ}$ & 1.55 ($0^{+}\rightarrow2^{+}$) & $68\pm8$ & 50.0 &
47.4\\\hline
\end{tabular}

\medskip

Table 1. Cross sections for Coulomb excitation of unstable nuclei.
The units of energy are MeV, the laboratory energy is in
MeV/nucleon, the B($\pi\lambda$)-values are in units of
e$^{2}$fm$^{2\lambda}$, and the cross sections are in millibarns. The
data for different experiments (numbered 1 to 12) were collected
from the references listed in column 1. The last two columns give
the calculated cross sections obtained by using eqs. \ref{cross_2}
and \ref{approx}, respectively.

\bigskip

\begin{figure}[ptb]
\begin{center}
\includegraphics[
height=3in, width=4in ]{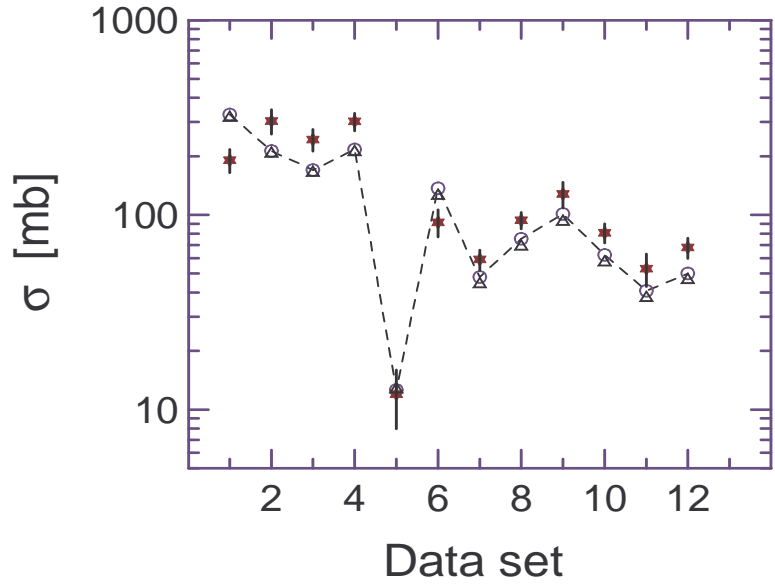}
\end{center}
\caption{Comparison  between experimental  Coulomb excitation cross sections (solid
stars with error bars) and theoretical  ones, calculated either with eq. \ref{cross_2} (open
circles), or with eq. \ref{approx} (open triangles).}%
\label{figcomp}%
\end{figure}

In figure \ref{figcomp} we show a comparison between the
experimental data and our calculations. We notice that the cross
sections calculated with help of eq. \ref{approx} are not much
different than those calculated with eq. \ref{cross_2}. They are
systematically lower, up to 10\%, than the exact calculation
following eq. \ref{cross_2}. As we discuss below, this is not always
the case, specially for the excitation of high-lying states.  In fact, this
is a good check of eq. \ref{cross_2}, which is done in a very
different way than the analytical calculations of eq. \ref{approx}.
But as we will see below, this agreement is not always the case,
specially when one includes small impact parameters for which the
sensitivity to the relativistic corrections is higher (see ref.
\cite{Ber03}). The
dashed curve in figure 1 is a guide to the eye. It helps to see that
the experimental cross sections are on average larger
than the calculated ones, either with eq. \ref{cross_2} (open
circles), or with eq. \ref{approx} (open triangles).

\begin{tabular}
[c]{|l|l|l|l|l|l|l|l|l|l|l|l|}\hline
& $E_{x}\ $[MeV] & $J_{i}^{\pi}\rightarrow J_{f}^{\pi}$ & $\pi\lambda$ &
$B(\pi\lambda)$ [e$^{2}$ fm$^{2\lambda}$] & 10 & 20 & 30 & 50 & 100 & 200 &
500\\\hline\hline
$^{11}$Be & 0.32 & $\frac{1}{2}^{-}\rightarrow\frac{1}{2}^{+}$ & E1 & 0.115 &
1128 & 653 & 473 & 315 & 187 & 115 & 69.6\\\hline
$^{11}$B & 2.21 & $\frac{3}{2}^{-}\rightarrow\frac{1}{2}^{-}$ & M1 &
2.40$\times10^{-2}$ & 0.301 & 0.799 & 1.15 & 1.63 & 2.33 & 3.08 & 4.17\\\hline
$^{11}$C & 2.00 & $\frac{3}{2}^{-}\rightarrow\frac{1}{2}^{-}$ & M1 &
1.52$\times10^{-2}$ & 0.196 & 0.551 & 0.793 & 1.12 & 1.57 & 2.07 &
2.76\\\hline
$^{12}$B & 0.953 & $1^{+}\rightarrow2^{+}$ & M1 & 4.62$\times10^{-3}$ &
0.227 & 0.395 & 0.490 & 0.607 & 0.762 & 0.917 & 1.13\\\hline
$^{12}$C & 4.44 & $0^{+}\rightarrow2^{+}$ & E2 & 37.9 & 34.6 & 38.6 & 31.3 &
21.6 & 12.1 & 6.93 & 3.81\\\hline
$^{13}$C & 3.09 & $\frac{1}{2}^{-}\rightarrow\frac{1}{2}^{+}$ & E1 &
1.39$\times10^{-2}$ & 8.37 & 11.3 & 11.0 & 9.61 & 7.28 & 5.39 & 3.89\\\hline
$^{13}$N & 2.37 & $\frac{1}{2}^{-}\rightarrow\frac{1}{2}^{+}$ & E1 &
3.56$\times10^{-2}$ & 38.2 & 43.6 & 39.6 & 32.5 & 23.2 & 16.4 & 11.4\\\hline
$^{15}$C & 0.74 & $\frac{1}{2}^{+}\rightarrow\frac{5}{2}^{+}$ & E2 & 2.90 &
8.79 & 4.04 & 2.65 & 1.59 & 0.839 & 0.475 & 0.267\\\hline
$^{16}$C & 1.77 & $0^{+}\rightarrow2^{+}$ & E2 & 2.12 & 8.81 & 4.41 & 2.92 &
1.76 & 0.920 & 0.517 & 0.285\\\hline
$^{16}$N & 0.12 & $0^{-}\rightarrow2^{-}$ & E2 & 10.2 & 31.0 & 14.1 & 9.21 &
5.53 & 2.91 & 1.64 & 0.926\\\hline
$^{17}$N & 1.37 & $\frac{1}{2}^{-}\rightarrow\frac{3}{2}^{-}$ & M1 &
5.15$\times10^{-3}$ & 0.153 & 0.304 & 0.397 & 0.516 & 0.680 & 0.848 &
1.09\\\hline
$^{17}$O & 0.87 & $\frac{5}{2}^{+}\rightarrow\frac{1}{2}^{+}$ & E2 & 2.07 &
6.30 & 2.88 & 1.87 & 1.12 & 0.588 & 0.332 & 0.184\\\hline
$^{17}$F & 0.5 & $\frac{5}{2}^{+}\rightarrow\frac{1}{2}^{+}$ & E2 & 21.6 &
68.3 & 29.7 & 19.3 & 11.6 & 6.08 & 3.44 & 1.92\\\hline
$^{18}$O & 1.98 & $0^{+}\rightarrow2^{+}$ & E2 & 44.8 & 109 & 60.7 & 40.9 &
24.8 & 11.6 & 7.27 & 3.99\\\hline
$^{18}$F & 0.94 & 1$^{+}\rightarrow3^{+}$ & E2 & 37.9 & 115 & 52.5 & 34.1 &
20.4 & 10.7 & 6.01 & 3.34\\\hline
$^{18}$Ne & 1.89 & $0^{+}\rightarrow2^{+}$ & E2 & 248 & 615 & 342 & 229 &
138 & 72.0 & 40.1 & 22.1\\\hline
$^{19}$O & 0.1 & $\frac{5}{2}^{+}\rightarrow\frac{3}{2}^{+}$ & M1 &
2.34$\times10^{-4}$ & 0.0495 & 0.0615 & 0.0673 & 0.0737 & 0.0799 & 0.779 &
0.799\\\hline
$^{19}$F & 0.11 & $\frac{1}{2}^{+}\rightarrow\frac{1}{2}^{-}$ & E1 &
5.51$\times10^{-4}$ & 8.07 & 4.36 & 3.06 & 1.97 & 1.10 & 0.592 & 0.337\\\hline
$^{19}$Ne & 0.24 & $\frac{1}{2}^{+}\rightarrow\frac{5}{2}^{+}$ & E2 & 119 &
361 & 157 & 102 & 61.6 & 32.5 & 18.5 & 10.5\\\hline
$^{20}$O & 1.67 & $0^{+}\rightarrow2^{+}$ & E2 & 28.0 & 72 & 37.4 & 24.9 &
15.1 & 7.86 & 4.41 & 2.43\\\hline
$^{20}$F & 0.656 & $2^{+}\rightarrow3^{+}$ & M1 & 3.56$\times10^{-3}$ &
0.237 & 0.385 & 0.465 & 0.560 & 0.683 & 0.803 & 0.959\\\hline
$^{20}$Ne & 1.63 & $0^{+}\rightarrow2^{+}$ & E2 & 319 & 834 & 433 & 287 &
173 & 89.8 & 50.3 & 27.6\\\hline
$^{30}$Ne & 0.791 & $0^{+}\rightarrow2^{+}$ & E2 & 460 & 1167 & 550 & 361 &
218 & 115 & 65.0 & 35.2\\\hline
$^{32}$Mg & 0.885 & $0^{+}\rightarrow2^{+}$ & E2 & 454 & 1151 & 541 & 355 &
214 & 112 & 63.0 & 36.7\\\hline
$^{42}$S & 0.89 & $0^{+}\rightarrow2^{+}$ & E2 & 397 & 945 & 445 & 292 & 175 &
91.9 & 52 & 29.7\\\hline
$^{46}$Ar & 1.55 & $0^{+}\rightarrow2^{+}$ & E2 & 190 & 399 & 209 & 140 &
84.4 & 44.1 & 24.7 & 13.6\\\hline
$^{54}$Ni & 1.40 & $0^{+}\rightarrow2^{+}$ & E2 & 626 & 1319 & 677 & 447 &
268 & 139 & 78.1 & 43.1\\\hline
\end{tabular}

\medskip

{Table 2 - Cross sections (in mb) for Coulomb excitation of
projectiles incident on Pb targets at bombarding energies ranging
from 10 to 500 MeV/nucleon. The energy units are MeV, the laboratory
energy is in MeV/nucleon, the B($\pi\lambda$)-values are in units of
e$^{2}$fm$^{2\lambda}$. }

\newpage

The cross sections for Coulomb excitation of numerous projectiles
incident on Pb targets at bombarding energies ranging from 10 to 500
MeV/nucleon are shown in Table 2.  These cross sections were
calculated assuming that the detectors collect events from all
possible Coulomb scattering events. In a real experimental
situation, the angular distribution is restricted to angular
windows, reducing the available cross sections. Only the lowest
lying transitions have been considered, i.e. from the ground to the
first excited states. One observes that some cross sections are very
large, specially for $^{11}$Be, $^{18}$Ne, $^{30}$Ne and $^{54}$Ni.
For these and other similar cases, the measurements are easy to
perform, with a large number of events/second even with modest
intensities. Cases such as $^{16}$C are well within the experimental
possibilities in most radioactive beam facilities.

Table 2 also shows that, except for M1 excitations, the Coulomb
excitation cross sections decrease steadily as the energy increases
from 10 to 500 MeV/nucleon. Based on these numbers alone, one could
conclude that Coulomb excitation of low-lying states (in contrast to
the case of high-lying states, e.g. giant resonances \cite{BB85})
are better suited for studies at low energies. However, reactions at
lower energies while are less influenced by contamination due to
nuclear breakup \cite{BCH95,Cha07} can give rise to large high-order
effects \cite{BC92}. The interpretation of data could be distorted
as in the case of Coulomb dissociation of $^8$B at low energy
\cite{Kolata}, which was completely misinterpreted in terms of
first-order calculations. In some situations, when higher-order
effects are relevant, the effect of the nuclear breakup cannot be
neglected either \cite{Gai,Vitturi}. Thus, the choice of the
incident energy would depend on the experimental conditions.
Identification of gamma-rays from de-excitation using Doppler shift
techniques are often more advantageous at higher energies. Moreover,
except for few cases (e.g. $^{11}$C), the magnetic dipole
transitions are much smaller than those for $E1$ and $E2$
transitions. Even for M1 transitions the measurements are under the
possibility of most new experimental facilities.

\bigskip

\begin{figure}[ptb]
\begin{center}
\includegraphics[
height=4.0906in, width=4.3154in ]{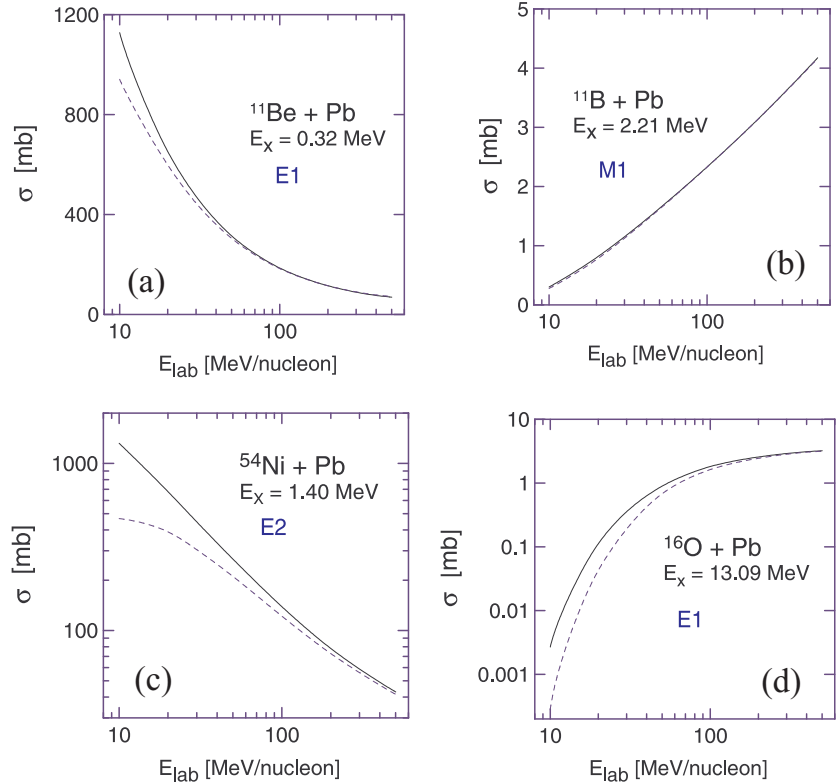}
\end{center}
\caption{Coulomb excitation cross section of \ the first excited state in
$^{11}$Be, $^{11}$B and $^{54}$Ni and of the 13.05 MeV sate in $^{16}$O
projectiles incident on Pb targets as a function of the laboratory energy.}%
\label{fig2}%
\end{figure}

\bigskip

The comparison of the exact calculations, using eq. \ref{cross_2}
(solid lines), and the approximations \ref{approx} (dashed lines)
are shown in figs. 2(a-d), for $^{11}$Be, $^{11}$B, $^{54}$Ni and
$^{16}$O, respectively. The $^{16}$O case (as well as for $^{12}$C
in Table 2) was included for comparison, with a high-lying excited
state. We see from figs 2a and 2b that the approximations in eq.
\ref{approx} work quite well for the $M1$ multipolarity and
reasonably well (within 20\% at 10 MeV/nucleon and 5\% at 50
MeV/nucleon) for the $E1$ cases. But they fail badly at low and
intermediate energies for the $E2$ ( fig. 2c). The reason is that
the $E2$ Coulomb field (\textquotedblleft tidal
field\textquotedblright) is very sensitive to the details of the
collision dynamics at low energies. These conclusions can be
deceiving since even for the $E1$ and $M1$ cases the approximations
in eq. \ref{approx} may strongly differ from the exact calculations
if the excitation energy is large (see discussion in ref.
\cite{Ber03}). This is shown in figure 2d, where we plot the Coulomb
excitation cross section of the $E_{x}=13.09$ MeV state in $^{16}$O.
In this case, the cross sections based on eq. \ref{approx} is a
factor of 10 smaller than the exact calculation at 10 MeV/nucleon.
At 100 MeV/nucleon this difference drops to 10\%, which still needs
to be considered with care.

In summary, in this article we have used the formalism of ref.
\cite{Ber03} to predict the cross sections for Coulomb excitation of
several light projectiles with electromagnetic transitions found in
the literature, listed in the TUNL database \cite{tunl}, and for a
few other selected cases. These estimates will be useful for planing
Coulomb excitation experiments at present and future heavy ion
facilities. It is evident that the inclusion of relativistic effects
combined with Coulomb distortion are of the utmost relevance. The cross section inferred by using
non-relativistic or pure relativistic treatments can be wrong by up
to 30\% even at 100 MeV/nucleon, as shown here and in ref.
\cite{Ber03}. Finally, the use of Coulomb excitation to produce
nuclei in high-lying states is an important tool to study particle
emission processes. For example, the excitation of $^{18}$Ne and its
subsequent decay by two-proton emission is a process of large
theoretical and experimental interest. Experimental work in this
direction is in progress \cite{Ra06}.

\section*{\bigskip}

\bigskip

\textbf{Acknowledgments}

This research was supported by the U.S. Department of Energy under contract
No. DE-AC05-00OR22725 (Oak Ridge National Laboratory) with UT-Battelle, LLC.,
and by DE-FC02-07ER41457 with the University of Washington (UNEDF, SciDAC-2).

\end{document}